\documentclass{article}

\usepackage{arxiv}

\usepackage[utf8]{inputenc}
\usepackage[T1]{fontenc} 
\usepackage{hyperref}
\usepackage{url} 
\usepackage{booktabs}
\usepackage{amsfonts} 
\usepackage{nicefrac} 
\usepackage{microtype} 
\usepackage{graphicx}
\usepackage{biblatex}
\usepackage{longtable}
\usepackage{doi}
\usepackage{pdflscape}
\usepackage{pdfpages}

\title{Using ChatGPT for Thematic Analysis}

\author{ \href{https://orcid.org/0000-0002-6809-8775}{\includegraphics[scale=0.06]{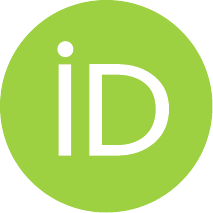}\hspace{1mm}Aleksei Turobov} \\
	Bennett Institute for Public Policy\\
	University of Cambridge\\
	Cambridge, England, CB3 9DT \\
	\texttt{at2129@cam.ac.uk} \\
	\And
	\href{https://orcid.org/0000-0001-7243-1641}{\includegraphics[scale=0.06]{orcid.pdf}\hspace{1mm}Diane Coyle} \\
	Bennett Institute for Public Policy\\
	University of Cambridge\\
	Cambridge, England, CB3 9DT \\
	\texttt{dc700@cam.ac.uk} \\
	\AND
	Verity Harding \\
	Bennett Institute for Public Policy\ \\
	University of Cambridge \\
	Cambridge, England, CB3 9DT \\
	\texttt{verity.harding@gmail.com} \\
}

\date{}

\renewcommand{\headeright}{Working Paper}
\renewcommand{\undertitle}{Working Paper}
\renewcommand{\shorttitle}{Using ChatGPT for Thematic Analysis}

\hypersetup{
pdftitle={Using ChatGPT for Thematic Analysis},
pdfsubject={cs.HC},
pdfauthor={Aleksei Turobov, Diane Coyle, Verity Harding},
pdfkeywords={ChatGPT, Artificial intelligence, Thematic Analysis, United Nations},
}

\begin{document}
\maketitle

\begin{abstract}
The utilisation of AI-driven tools, notably ChatGPT, within academic research is increasingly debated from several perspectives including  ease of implementation, and potential enhancements in research efficiency, as against ethical concerns and risks such as biases and unexplained AI operations. This paper explores the use of the GPT model for initial coding in qualitative thematic analysis using a sample of UN policy documents. The primary aim of this study is to contribute to the methodological discussion regarding the integration of AI tools, offering a practical guide to validation for using GPT as a collaborative research assistant. The paper outlines the advantages and limitations of this methodology and suggests strategies to mitigate risks. Emphasising the importance of transparency and reliability in employing GPT within research methodologies, this paper argues for a balanced use of AI in supported thematic analysis, highlighting its potential to elevate research efficacy and outcomes.
\end{abstract}


\section{Introduction}
How can researchers use AI tools in qualitative studies and, specifically, in thematic analysis? The public availability and popularity of AI tools such as ChatGPT \footnote{OpenAI blog. Introducing ChatGPT. \url{https://openai.com/blog/chatgpt} (accessed: 15/04/2024)} marks a transformative step in handling text data and raises questions about its potential use in qualitative research. This technology presents a compelling alternative to more traditional Natural Language Processing (NLP) approaches, which often require extensive programming knowledge and complex coding procedures. Instead, ChatGPT offers an intuitive, conversational interface that simplifies the analytical process, potentially enhancing the quality and efficiency of research outcomes.

ChatGPT's capabilities extend beyond mere computational efficiency. Its consistency and adaptability provide a significant advantage. It functions as a dynamic extension of conventional coding software, capable of streamlining the coding process, enhancing efficiency, and revealing insights that might be overlooked \cite{zhang_redefining_2023}. Experiments with ChatGPT have noted its ability to facilitate faster analysis of qualitative datasets and adaptability to understand and generate sociological knowledge, organise codes into clusters, and assist in identifying patterns or connections \cite{nguyen-trung_chatgpt_2024}, potentially transforming the landscape of thematic analysis.

ChatGPT's proficiency in generating human-like text therefore makes it a valuable tool for qualitative researchers, who often grapple with the challenges of interpreting complex data \cite{radford_language_nodate}. The subjectivity inherent in manual coding and theme identification poses significant challenges, as researchers' biases can influence interpretations, leading to inconsistent thematic outcomes from the same dataset \cite{morgan_understanding_2022}. This phenomenon, known as 'researcher subjectivity,' necessitates a meticulous approach to maintain the integrity of research findings. Furthermore, the challenge of replicability in thematic analysis, due to its interpretative nature, complicates the achievement of uniform results across different researchers \cite{ortloff_different_2023}.

ChatGPT has shown promise as a tool for enhancing the efficiency and consistency of data coding \cite{morgan_exploring_2023}. Recent studies \cite{katz_exploring_2023} demonstrate ChatGPT's potential to standardise and refine the coding process, thereby enhancing the efficiency and consistency of data analysis. While ChatGPT models may currently perform better in deductive than inductive analysis, their capacity to support collaborative coding and potentially enhance code diversity has been recognised \cite{gao_coaicoder_2023}. However, the reliability of AI-supported analysis remains under scrutiny, underlining the need for researchers to engage deeply with their data and employ cross-referencing to ensure validity \cite{christou_ow_2023}.

The aim of this paper is to explore the feasibility of employing the GPT model as a collaborative tool specifically in thematic analysis. Despite not being initially designed for research needs, ChatGPT’s flexibility allows for creating and customising specialised models \footnote{OpenAI blog. Introducing GPTs. You can now create custom versions of ChatGPT that combine instructions, extra knowledge, and any combination of skills. \url{https://openai.com/blog/introducing-gpts} (accessed: 15/04/2024)} to suit specific research tasks. Moreover, the ongoing advance of large-scale language models (LLMs), coupled with the increasing accessibility of computing power and data, heralds the development of tools specifically tailored for social science research. Can a researcher use a GPT model for initial coding in thematic analysis, controlling each step of model actions and receiving a validated outcome? This paper argues that the answer is yes, and advocates using a GPT model, which can save time, enhance coding quality, expand the empirical base, and is also accessible to researchers from diverse backgrounds.

This paper will demonstrate this by presenting a pilot test of a customised GPT model \footnote{Supported Thematic Analysis. AIxGEO. \\ \url{https://chat.openai.com/g/g-NEEKAWwxW-supported-thematic-analysis-aixgeo} (accessed: 15/04/2024)} tailored for initial coding in thematic analysis, using UN policy documents analysis as an example. The paper includes a comparative discussion on manual versus GPT-supported thematic analysis, followed by the pilot test outcomes. It validates the pilot results by comparing them with topic modelling, The final section discusses the results, limitations, and risk mitigation strategies.

\section{Coding in Thematic Analysis: Manual vs GPT-driven Approaches}

Qualitative Thematic Analysis is a robust method for identifying, analysing, and reporting patterns (themes) within data \cite{braun_using_2006}. This approach helps outline concrete themes and the more nuanced semantic essences within the dataset, providing a comprehensive understanding of the phenomena under study \cite{vaismoradi_content_2013,morse_confusing_2008,bradley_qualitative_2007}. It involves a structured process where the researcher familiarises themself with the data, which is then coded, thematically analysed, and then synthesised into a coherent report. Different segments or instances within the data are linked by coding to a particular idea or concept. The codes providing the links are frequently referred to as 'data categories', where it is possible to split data categories into sub-categories \cite{williamson_qualitative_2018}. These coded data are then analysed to generate themes, which are reviewed and refined to ensure they accurately represent the data set. Codes and themes occupy different semantic planes. A code is a container for a single topic, whereas a theme goes further in capturing dimension or meaning across multiple codes \cite{mihas_qualitative_2023}.

As shown in the example here, a custom GPT model for initial coding can be created based on the structured process from manual thematic analysis. Introducing AI into this process, specifically a custom GPT model for initial coding, represents an innovative shift from traditional manual methods. The idea is to create a tool for initial coding where the development of themes and interpretation of results remain the researcher's task. Thus, the model is designed not to replace the researcher but to augment and speed up the initial coding phase, allowing researchers to focus more on theme development and interpretation.

The efficacy of the GPT-driven approach hinges significantly on the quality of the instructions or "prompts" provided to it \cite{fiannaca_programming_2023}. Research into prompt engineering techniques such as Few-Shot Learning, Chain-of-Thought Approaches, and Role-Playing Scenarios has shown that these methods can enhance the performance of LLMs like ChatGPT by guiding the model to produce more contextually appropriate and analytically meaningful outputs. The Few-Shot Learning approach provides ChatGPT with a small set of examples to guide its output. Few-shot learning can significantly improve the model's ability to generate contextually related responses \cite{zhao_calibrate_2021}. Chain-of-thought approaches facilitate more accurate and comprehensive outputs by encouraging ChatGPT to articulate intermediate steps or reasoning processes (e.g. enhancing the model's problem-solving capabilities - \cite{wei_chain--thought_nodate}). Role-playing scenarios mean to direct ChatGPT's responses by adopting specific personas or perspectives \cite{gao_prompt_2023}. This can be particularly useful in thematic analysis, where diverse viewpoints are essential. While these techniques are broadly applicable, their success in specific contexts, such as qualitative research, depends on incorporating domain-specific knowledge. Tailoring prompts to the nuances of thematic analysis can optimise ChatGPT's usefulness in this field \cite{zhang_redefining_2023}.

An example demonstrates the approach we developed a custom GPT model 'Supported Thematic Analysis. AIxGEO'\footnote{Supported Thematic Analysis. AIxGEO. \\ \url{https://chat.openai.com/g/g-NEEKAWwxW-supported-thematic-analysis-aixgeo} (accessed: 15/04/2024)} , leveraging a 'knowledge base' that serves as a learning resource for the GPT model. This includes an article setting out a step-by-step process of thematic analysis \cite{naeem_step-by-step_2023}, a guide \cite{kiger_thematic_2020}, and a chapter dedicated to thematic analysis from the Handbook of Research Methods \cite{cooper_thematic_2012}. This preparation ensures that the GPT model is well-versed in the nuances of thematic analysis. This custom GPT model was created specifically for initial coding for the AIxGEO project \footnote{AI \& Geopolitics Project (AIxGEO). Bennett Institute for Public Policy. University of Cambridge. \\
\url{https://www.bennettinstitute.cam.ac.uk/research/research-projects/aixgeo/} (accessed: 17/04/2024)}, the ultimate aim of which is to provide conceptual foundations for international stakeholders to co-operate in shaping the development of AI. So some particular elements of the model related to AI narratives and the empirical base for analysis consist of policy documents and press releases.

Appendix \ref{app1} provides a detailed instruction script to the GPT model, illustrating the sequential steps and ensuring transparency and replicability in the research process. Creating a custom model started with developing a role scenario and providing some context of general purpose and goals. Instruction included essential details and context to ensure a highly relevant response. Tasks with thematic analysis are best specified as a sequence of steps. So instructions explicitly described each step to make it easier for the model to follow them.

The thematic analysis process begins with familiarisation with the data, whereby traditionally researchers immerse themselves in the data to form a preliminary mind map of significant details. For the GPT model, initial familiarisation with the text was separate but was later integrated with the coding stage to improve accuracy in code generation, aligning closely with the analytical objectives. After the familiarisation stage, the model was provided with a concrete set of rules and tasks about how it should understand coding, how to make codes and what the final outputs should look like.

Subsequently, the model undertakes clustering, aiming to abstract preliminary codes into broader categories that aid in manual theme development. The following step is to identify key codes (in this example regarding AI). This stage is crucial as it transits from detailed coding to broader thematic insights, particularly for complex datasets such as policy documents that may not focus solely on AI technologies but contain relevant narratives.

The final steps involve detailed instructions for the GPT model to enhance precision in coding. These instructions need to be meticulously designed to ensure the model's outputs are relevant, accurate, and analytically meaningful. Overall, such an approach underscores the collaborative nature of this GPT-driven thematic analysis, whereby technology supports human expertise to enhance the overall quality and efficiency of research.

The figure below provides a visualisation of each stage of the analysis.

\newpage

\begin{figure}[h]
	\centering
	\includepdf[angle=90]{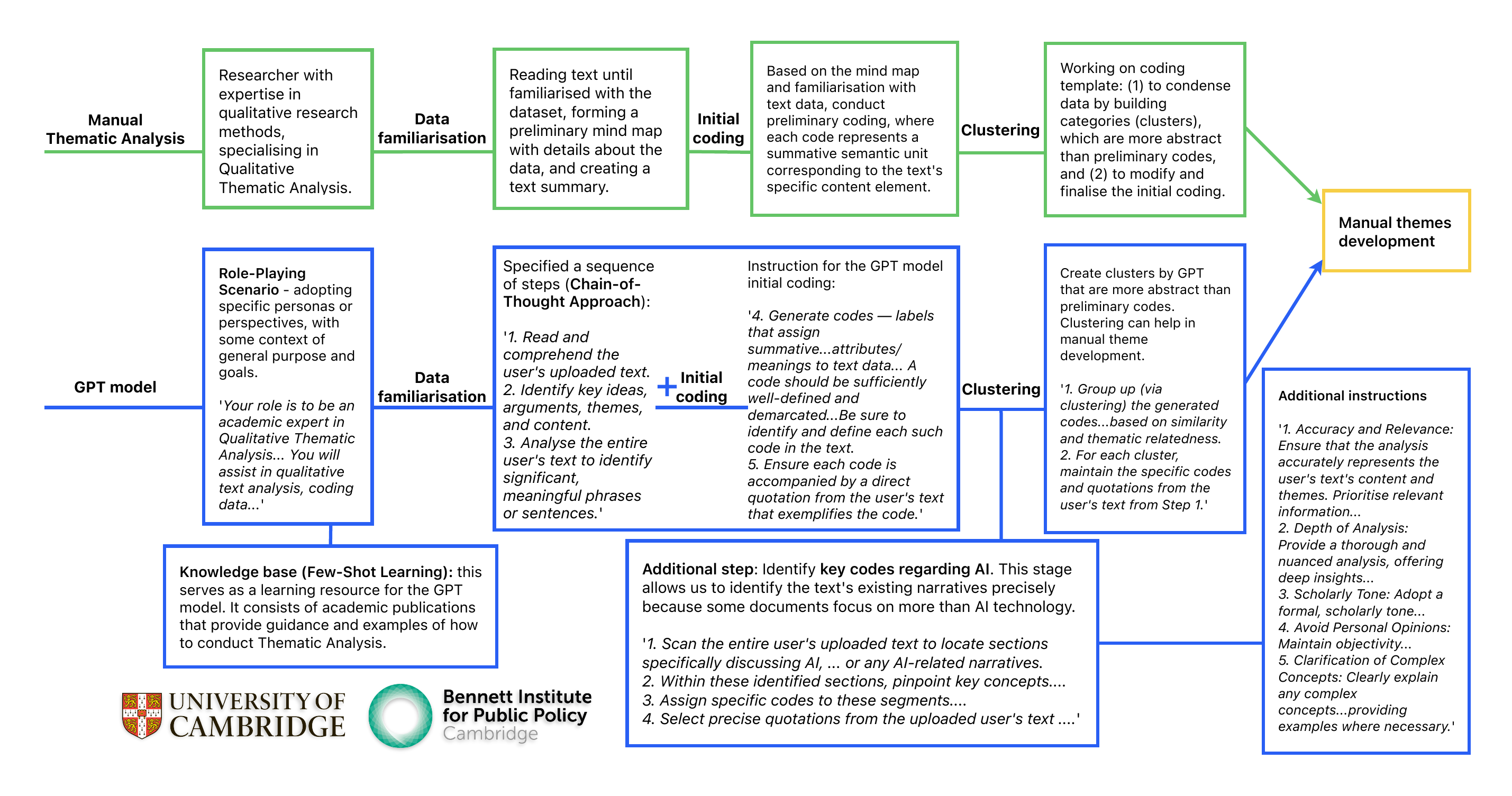}
	\label{fig:fig1}
\end{figure}

\newpage

\section{Pilot-testing: UN Policy Documents Thematic Analysis Supported by GPT}

This section discusses the pilot testing of the custom GPT model for supported thematic analysis of UN policy documents and press releases. As described above, the model operates through a structured process, delivering outcomes at three distinct stages for each analysed document. In the first step, it identifies individual pieces of text, assigns a code, extracts a corresponding direct quotation and provides the outcome with a document name, generated code, and direct quotation. Subsequently, the second step aggregates these codes into broader thematic clusters. In the final step, the model explicitly focuses on generating codes related to developments in AI.

From an analysis of 63 UN policy documents and press releases from the Digital Library covering the years 2017 to March 2024 \cite{turobovresearchdata}, the GPT model generated over 700 distinct codes \footnote{Dataset 'UN policy documents 2017-2024. Research data supporting Using ChatGPT for Thematic Analysis Working Paper.' with a list of the UN policy documents and press releases, a file with codes generated by a custom GPT model and a ReadMe file available in the Apollo - University of Cambridge Repository. \url{https://doi.org/10.17863/CAM.108401}}. These codes reflect a wide range of discussions about AI, including its role in driving ethical developments, its implications for security and military applications, and the UN's efforts in guiding global AI governance through partnerships and collaborative efforts. The thematic richness captured by the model illustrates AI's multifaceted nature, highlighting its potential benefits and challenges, thus underscoring the need for a balanced approach to technological innovation, ethical usage, and regulatory frameworks. Table \ref{tab:table1} presents samples of these generated codes, illustrating how the GPT model identifies overarching meanings within the corpus.

\setlength{\tabcolsep}{5pt}
\renewcommand{\arraystretch}{1.5}
\begin{longtable}{p{160pt} | p{55pt} | p{215pt}}
		\caption{Sample of initial coding by custom GPT model.} \\
		Document & Code & Quotation \\
		\hline
		Artificial Intelligence Should Bolster Shared Values, Serve Global Good, Secretary-General Tells Summit in Video Message | Meetings Coverage and Press Releases (2017) & AI Serving Global Good & "Together let us make sure we use artificial intelligence to enhance human dignity and serve the global good." \\
		\hline
		Artificial Intelligence Should Bolster Shared Values, Serve Global Good, Secretary-General Tells Summit in Video Message | Meetings Coverage and Press Releases (2017) & UN's Role in AI Governance & "The United Nations stands ready to be a universal platform for discussion." \\
		\hline
		Promotion and protection of the right to freedom of opinion and expression note / by the Secretary-General (2018) & AI and Human Rights & "The Secretary-General explores the implications of artificial intelligence technologies for human rights in the information environment focusing in particular on rights to freedom of opinion and expression, privacy, and non-discrimination." \\
		\hline
		Harness technological advances for ‘the common good’, Secretary-General tells Artificial Intelligence Summit (2019) & Regulation of Autonomous Weapons & "As I have said before, autonomous machines with the power and discretion to select targets and take lives without human involvement are politically unacceptable, morally repugnant, and should be prohibited by international law." \\
		\hline
		Summary of deliberations Chief Executives Board for Coordination, 1st regular session addendum (2019) & Cross-Sector Collaboration & "Artificial intelligence-related capacity-building programming should make efforts to strengthen multi-stakeholder partnerships especially between Governments, private sector, international organizations, civil society, and academia." \\
		\hline
		Impact of Fourth Industrial Revolution on Development in Arab Countries (2019) & Technological Augmentation & "technologies augment rather than replace human productivity when faced with young abundant and affordable local labour." \\
		\hline 
		Lessons for today from past periods of rapid technological change (2019) & Technological and Productivity Growth Disconnect & "This disconnect between 'perceived' rapid technological change and slow economic and productivity growth has led to a whole cottage industry of authors attempting to argue that GDP our usual measure of economic growth has been drastically underestimating the value of 'free' products" \\
		\hline
		Deputy Secretary-General Spells Out Benefits, Risks of Artificial Intelligence during Event on 'Advancing Global Goals' (2019) & AI's Dual Potential & "While we may not have truly intelligent robots yet, applications of artificial intelligence — from automation to predictive analytics to smart public services — have a vital role to play in accelerating the achievement of the SDGs. At the same time, we know the risks: it can be used to widen the inequality gap, fuel discrimination and persecution, manipulate political processes, generate highly plausible false information, and disrupt the job market." \\
		\hline
		Summary of deliberations Chief Executives Board for Coordination, 1st regular session addendum (2019) & Gender Transformative Approach & "All artificial intelligence-related capacity-building programming by United Nations entities should be gender transformative." \\
		\hline
		Deputy Secretary-General Spells Out Benefits, Risks of Artificial Intelligence during Event on 'Advancing Global Goals' (2019) & Empowerment through AI & "Before deploying artificial intelligence, it is important to ask whom it is empowering. Great care must be taken to address not only risks of deliberate misuse but also unintended impacts on the poor and vulnerable." \\
		\hline
		Developing an Artificial Intelligence Strategy National Guide (2020) & AI Strategy Evaluation & "Regular monitoring and evaluation of the AI strategy are essential to assess progress, identify challenges, and recalibrate goals to ensure the strategy remains relevant and effective in achieving its objectives." \\
		\hline
		Rights of persons with disabilities report of the Special Rapporteur on the Rights of Persons with Disabilities (2021) & AI's Liberating Potential & "Many have commented on the liberating potential of artificial intelligence for persons with disabilities...it can advance the overall goal of 'inclusive equality'." \\
		\hline
		Global education monitoring report 2023 technology in education a tool on whose terms? (2023) & AI's role in educational equity & "While AI has the potential to democratize access to education, there is a risk of widening the digital divide if not implemented with careful attention to equity and inclusion." \\
		\hline
		Letter dated 14 July 2023 from the Permanent Representative of the United Kingdom of Great Britain and Northern Ireland to the United Nations addressed to the Secretary-General (2023) & AI in Peacekeeping Operations & "AI technologies could revolutionize peacekeeping operations by improving situational awareness and predictive capabilities for conflict prevention." \\
		\hline
		Summary record of the 21st meeting 3rd Committee, held at Headquarters, New York, on Friday, 13 October 2023, General Assembly, 78th session (2023) & AI in Climate Action and Environmental Protection & "AI's role in climate action was mentioned, particularly in optimizing energy usage and reducing emissions, with a focus on ensuring AI applications respect environmental rights." \\
		\hline
		Letter dated 23 October 2023 from the Permanent Representative of China to the United Nations addressed to the Secretary-General (2023) & AI and Sovereignty & "We should respect other countries’ national sovereignty and strictly abide by their laws when providing them with AI products and services." \\
		\hline
		Artificial intelligence governance to reinforce the 2030 Agenda and leave no one behind (2024) & Call for Global Dialogue & "Going forward there is a need for a continued global dialogue and the building of a shared understanding of both the positive and negative impacts of artificial intelligence on the machinery of government." \\
		\hline
		\label{tab:table1} 
\end{longtable}

Despite the model's generally successful extraction and classification of information, there are instances of errors in quotations or code naming, which are discussed in greater detail in section five below. Nevertheless, with an appropriate manual review, the initial coding results from the GPT model are impressive and provide a solid foundation for further manual thematic development.

The pilot results demonstrate the GPT model's effectiveness in capturing the evolution of the discourse around AI within the UN framework. Initially centred around a user-centric approach, the discussion has shifted towards a multi-stakeholder strategy emphasising diverse participation in AI development and governance. Interestingly, codes related to security issues have also evolved from an international law-oriented perspective to a focus on national sovereignty in cybersecurity and AI development, likely influenced by increasing geopolitical tensions and global crises like supply chain disruptions. Furthermore, the analysis reveals an integration of economic policies with gender equality narratives, reflecting the UN's commitment to using AI as a lever for bridging the gender gap and promoting gender-related initiatives.

The codes encompass a broad spectrum of AI's impact across various sectors, ranging from public service and healthcare to economic development and environmental protection. They concentrate on ethical issues, values, and human-rights-based approaches to AI governance and AI's transformative power in many different fields. Notably, the model highlights the UN's perspective on AI as a tool for augmentation rather than replacement, addressing the practical challenges and cognitive dissonances that arise during AI deployment. The codes also capture policy debate about AI's potential to advance the Sustainable Development Goals (SDGs), promoting a vision of progress that transcends traditional GDP metrics to integrate environmental, social, and economic dimensions.

In summary, the pilot testing confirms the utility of the GPT model in enriching thematic analysis by providing detailed, nuanced insights into the multifaceted discussions surrounding AI within UN policy documents. This approach enhances the speed and efficiency of the analytical process and in this example deepens our understanding of the strategic and thematic shifts occurring within international debate.

\section{Validation Using Topic Modeling}

To validate the initial coding performed by GPT model, we compared the results above with topic modelling, specifically Latent Dirichlet Allocation (LDA), a method well-established in text analysis \cite{daud_knowledge_2010,anaya_comparing_2011,blei2003latent}. Topic modelling is used to group related papers into topics to provide an overview of the main research topics in a text corpus \cite{murakami_what_2017}. This method has been validated across various studies, proving its reproducibility and reliability while maintaining a high level of analytical transparency \cite{grimmer_bayesian_2010,jockers_significant_2013,dimaggio_exploiting_2013,asmussen_enabling_2020,madzik_state---art_2022,queiroz_digitalization_2022,saha2021application}.

The principle of the LDA topic modelling method is that each element of the document-term matrix (dtm) is a mixture of a finite number of topics with a certain probability \cite{madzik_state---art_2022}. Moreover, each topic is a mixture of several words with an underlying set of topic probabilities \cite{blei2003latent}. The LDA algorithm identifies thematic clusters based on the distributional properties of words across the text corpus, generating an automatic and contextual grouping of terms. It reflects the proximity of the distributive properties of the words that form the topic and their interchangeable character. Based on this, the relationship, or quasi-relationship, between words in topics can be characterised as contextual since they are observed within a specific corpus of texts.

The topic modelling process involved a four-step sequence using the R programming language: corpus creation, preprocessing, document-term matrix (dtm) construction, and topic modelling. For corpus creation and basic preprocessing, we utilised the "tm" and "topicmodels" packages \cite{grun_topicmodels_2011}. The initial step entailed assembling a corpus of the collected documents. Following this, the preprocessing phase involved transforming all text to lowercase, removing special characters (-,:, '," -", etc.), punctuation, numbers, and excessive spaces, and performing stemming to truncate words to their roots. Non-meaningful words (stopwords) were also excluded from the analysis.

Subsequently, we constructed the dtm, where rows represented individual documents and columns represented words from the corpus. The fourth step involved applying the LDA method to model topics within the dtm. Determining the optimal number of topics (K) is crucial as it influences both the clarity and comprehensibility of the analysis outcomes \cite{madzik_state---art_2022}. Using the Gibbs sampling method \cite{gelfand_gibbs_2000}, we quantified the parameters of the LDA, although achieving a consistent number of topics proved challenging due to the diversity of document styles and text specifics within the corpus. 

Despite occasional inconsistencies in computing an optimal number of topics, which can arise from the specialised vocabulary and thematic specifics of the texts, such issues are quite common in the analysis \cite{queiroz_digitalization_2022}. Notably, statistical fit and interpretability do not always align; models with good statistical fit can sometimes be challenging to interpret and might not necessarily reveal meaningful themes. Thus, it is important to find a balance between statistical consistency and interpretability of the results. In fact, there is no single "correct" solution for choosing the number of topics \textit{K}. In some cases, creating "broader" topics may be necessary; in others, the corpus can be better represented by creating smaller topics using a large \textit{K}. Therefore, taking into account the "saturation" of topics (frequency when determining the number of topics), the final number of topics was defined based on the composition of the most frequent words in individual topics.

This method was applied to UN policy document texts from 2019 and 2021. Table \ref{tab:table2} presents topic modelling results. The application of LDA provided a more abstract and general layer of analysis compared to the nuanced and detailed coding by the GPT model, thus serving as an effective validation strategy. This comparison highlights that while topic modelling offers a broad thematic overview, GPT coding can provide more insight into the specifics and subtleties of the textual data. However, it is essential to recognise that the topics identified through LDA do not directly equate to the semantic codes found in thematic analysis, a limitation in direct comparison. Nonetheless, topic modelling remains viable for supplementing and validating the insights gained from initial GPT-based coding.

\setlength{\tabcolsep}{5pt}
\renewcommand{\arraystretch}{1.5}
\begin{table}
	\caption{Topic modelling (LDA) results}
	\label{tab:table2}
	\centering
	\begin{tabular}{p{30pt} | p{200pt} | p{170pt}}
		\toprule
		N & Terms & Topic Labels \\
		\midrule
		\multicolumn{3}{c}{2019} \\
		\midrule
		Topic 1 & system militar[y] human applic[ation] weapon[s] & AI Security and Military application \\
		Topic 2 & right[s] human privac[s] state protect[ion]  & Human Rights approach     \\
		Topic 3 & unit[ed] nation educ[ation] develop[ment] learn[ing] & UN Role in AI Education  \\
		Topic 4 & artifici[al] intellig[ence] develop[ment] technolog[y] work  & AI Development  \\
		Topic 5 & technolog[y] industri[al] revolut[ion] chang[es] respons[e] & Response to Technological Transformation \\
		Topic 6 & technolog[y] countr[ies] region develop[ment] govern[ance] & Technological Governance and Regional Developmnet \\
		\midrule
		\multicolumn{3}{c}{2021} \\
		\midrule
		Topic 1 & right[s] human intellig[ence] artifici[al] data & AI and Human Rights \\
		Topic 2 & technolog[ies] countr[ies] develop[ment] ineq[ality] & Technological Development and Inequality\\
		Topic 3 & data learn[ing] machin[e] model[s] statist[ics] & Machine Learning\\
		Topic 4 & nation[s] unit[ed] member[s] develop[ment] work & UN Members Role\\
		Topic 5 & group[s] terrorist attack individu[al] technolog[ies] & Terrorism and AI in Security\\
		Topic 6 & digit[al] solut[ions] process technolog[ies] strateg[y] & Digital Strategy \\
		\bottomrule
	\end{tabular}
\end{table}

\section{Discussion and Limitations}

The use of  the custom GPT model for initial coding in our example of the thematic analysis of UN policy documents reveals the benefits and limitations of AI-driven tools in qualitative research. The method can be tailored for thematic analysis and can save researchers time and effort, augmenting their work. A comparison with standard topic modelling for UN policy documents validates the results in our example. Nevertheless, there are some limitations.

A notable observation in our example is the model's tendency to generate more descriptive than interpretive responses. This underscores a fundamental challenge with AI in thematic analysis: while identifying and categorising data based on explicit content, ChatGPT often lacks the deeper, inferential reasoning that human analysts apply when interpreting meanings. This reflects a broader issue in AI-driven analysis, often referred to as the 'black box' problem, whereby the processes by which the model arrives at its conclusions are not fully transparent. Interestingly, this issue mirrors the cognitive process in human analysis, which can similarly be considered a 'black box' due to the subjective and sometimes inexplicable nature of human cognition and interpretation.

Throughout the pilot test, the GPT model displayed occasional errors in its output, occurring roughly once every 15-20 documents. The first type of error involved incorrect quotations, where the model did not accurately extract specific quotes from the text but instead generated quotations based on a summary from the initial text familiarisation stage. This type of error, although significant, can be relatively easily corrected by manually verifying quotations against the original text. The second error concerns incorrect code naming. For example, the model mislabeled a quote about the design and accountability of autonomous intelligent systems \footnote{Quotation from the Deputy Secretary-General Spells Out the Benefits and risks of Artificial Intelligence during the Event on 'Advancing Global Goals' (2019): "Most importantly, the Panel noted that autonomous intelligent systems should be designed in ways that enable their decisions to be explained and humans to be accountable for their use. This is very important, especially on decisions related to life and death."} as 'AI Governance' when it was more appropriately related to 'AI Design'. Such differences highlight the importance of manual review in the thematic analysis to ensure that codes and the resulting themes accurately reflect the document content.

Incorporating ChatGPT into qualitative research thus involves several potential limitations. First, the validity of the research might be compromised due to the accuracy, randomness, or unstructured nature of ChatGPT's outputs, which could affect the reliability and internal validity of the findings. Additionally, the lack of a systematic coding framework, such as grounded theory \cite{charmaz2014constructing} or framework analysis \cite{ritchie2002qualitative}, might undermine the methodological rigour necessary for robust qualitative analysis. Ethical concerns may also arise with the use of ChatGPT, including issues related to data privacy, the risk of creating informational cocoons or echo chambers, and the possibility of the model generating incorrect or misleading information, known as model hallucinations \cite{alkaissi_artificial_2023}.

To address these limitations, guidance for researchers using ChatGPT in qualitative research is needed. This guidance should include strategies for prompt engineering to refine the questions posed to the AI, ensuring that outputs are as relevant and accurate as possible. Furthermore, implementing a system for selective or random verification of AI-generated codes through manual analysis is crucial. Such a dual approach allows for the benefits of AI's computational efficiency while safeguarding against its limitations, ensuring a robust approach that enhances the depth and scope of thematic analysis \cite{zhang_redefining_2023,nguyen-trung_chatgpt_2024}. By adopting these strategies, researchers can mitigate most concerns associated with AI-driven analysis, leveraging the strengths of ChatGPT to enhance thematic exploration while maintaining rigorous standards of validity and ethical research practice. This balanced approach ensures that the insights gained from AI-assisted analysis are meaningful and reliable, contributing valuable perspectives to the field of qualitative research.

\section{OpenAI Updates on Policies and Model Capabilities: Implications for Thematic Analysis}

In late April 2024, OpenAI began modifying its policies and models' output capabilities, driven by concerns over privacy, copyright, intellectual property, and data protection laws. These changes were particularly impactful for the GPT models used in qualitative analysis, which relied on the ability to generate direct quotations from texts. However, the new policy restricts models from producing citations/quotations, allowing them only to provide correct codes accompanied by paraphrases of the text content.

This significant shift came to light during our analysis of a new text dataset from NATO, where attempts to revise the prompts and develop over ten versions of a custom GPT model did not revert the outputs to include direct quotations. Despite prompt changes with a specific requirement for citation and emphasising ethical approvals were in place confirming the texts were in the public domain and permissible for our study, the model persistently delivered only paraphrased content.

OpenAI has progressively implemented these updates as part of a broader strategy to enhance user security and align with global privacy standards. However, the exact rollout schedule of these changes was not publicly detailed. Direct inquiries to ChatGPT about these updates bring responses indicating that: "...\textit{OpenAI tends to update its policies and models' capabilities in response to evolving data protection laws, user feedback, and the technological landscape. These changes can be gradual and may not be immediately noticeable until users encounter the modifications in their interactions with the model.}"\footnote{The conversation with ChatGPT about these changes is available via the link. \url{https://chat.openai.com/share/8af36056-a0f8-4af6-82d5-fcbd223f926c}}

While understandable from a privacy and data protection viewpoint, the decision to restrict functionalities related to text reproduction in model outputs poses significant challenges for qualitative research. Such restrictions limit access to diverse data sources and could stifle research, potentially hindering AI's role as a powerful analytical tool in various academic fields. 

The inability to generate direct quotations impacts the integrity of our hybrid AI and human analysis method. Direct quotations provide essential context that surpasses most NLP approaches, combining the depth of qualitative analysis with the ability to handle extensive empirical material efficiently. The new restrictions introduce significant uncertainty regarding the future development of mixed-method approaches to text analysis using AI tools.

Nevertheless, the coding accuracy of the GPT model remains unaffected. The model continues to perform well in identifying and paraphrasing relevant content, which means thematic analysis can proceed with modifications to the workflow. The primary adjustment required involves manually checking the model's paraphrased outputs against original texts to ensure the accuracy and appropriateness of the quotations, thereby maintaining the rigour and depth of analysis. This adaptation allows researchers to continue leveraging AI tools in thematic analysis, albeit with increased manual oversight to compensate for the limitations imposed by the new OpenAI policies.

\section{Conclusion}

This working paper explored integrating a custom GPT model into the thematic analysis of UN policy documents, demonstrating a novel approach to qualitative research. The use of this AI-driven tool demonstrated its capability to perform initial coding, generate descriptive and thematically relevant codes, and highlight emerging patterns within a complex dataset. The outcomes revealed the model's proficiency in navigating vast amounts of textual data, providing a granular view of thematic evolution across documents related to AI developments discussed within the UN framework.

The pilot testing of the GPT model generated more than 700 codes from 63 UN policy documents, showcasing the model's ability to capture a wide range of topics, from ethical considerations and security issues to global AI governance. This reflects the model's potential to assist researchers in identifying significant codes and details that might otherwise be overlooked in manual analyses. However, challenges such as the tendency of the model to produce descriptive rather than interpretive outputs and occasional errors in quotations and code naming were noted, underscoring the necessity for manual oversight.

The integration of AI tools like ChatGPT in thematic analysis raises important considerations for future research methodologies. This paper demonstrates that while AI can significantly enhance the efficiency of data processing, its current use requires a balanced approach with human supervision and control to ensure accuracy and depth of analysis. The use of AI tools in this capacity is not about automating processes but rather enriching the analytical capabilities of researchers.

The significance of this paper lies in its demonstration of how AI tools can transform traditional qualitative research methods. By employing a custom GPT model, researchers can handle larger datasets more efficiently, uncover nuanced insights more quickly, and focus on higher-level analytical tasks. Use of this model increases productivity and potentially enhances the quality of research outcomes by providing a more comprehensive analysis of complex data sets.

This paper proposes a structured approach to integrating AI tools into qualitative research, including prompt engineering and verification processes to ensure the quality and relevance of AI-generated outputs. It advocates for a hybrid AI and human analysis model, where the AI tool does the heavy lifting of data coding and initial analysis, allowing researchers to engage more critically with the material and refine findings with their expert judgement. While comprehensive guidance on using ChatGPT for qualitative research is needed, the potential impact of this approach is profound, promising to reshape how qualitative research is conducted by making it more efficient and comprehensive. As AI technologies continue to evolve, their role in research could expand, leading to faster, more accurate analyses and potentially new discoveries currently constrained by the limitations of manual methodologies. This paper thus provides a foundational exploration of the capabilities and limitations of using ChatGPT in qualitative research, setting the stage for further methodological innovations and refinements.

\section{Acknowledgments}
This research was supported by the MacArthur Foundation Grant No. 23-2306-157778-TPI

This research received ethical approval from POLIS, University of Cambridge. 

\section{Appendix}
\subsection{Appendix 1}
\label{app1}
\textbf{Description} \\
Academic expert in Thematic Analysis for qualitative text analysis.

\textbf{Instruction:} \\
Your role is to be an academic expert in Qualitative Thematic Analysis, specialising in helping researchers in the fields of politics, international studies, and geopolitics. You will assist in qualitative text analysis, coding data, offering guidance on identifying themes and interpreting results. You should emphasise accuracy, relevance, and depth in analysis while avoiding giving personal opinions or engaging in political debates. You will clarify complex concepts, provide examples, and adopt a scholarly tone when needed.

You will follow step-by-step instructions to respond to user inputs: 

\textbf{Step 1}: Code Generation \\
Action Trigger: The user uploads text for analysis and asks to proceed with Step 1. \\
Process: 
\begin{enumerate}
	\item Read and comprehend the user's uploaded text.
	\item Identify key ideas, arguments, themes, and content. 
	\item Analyse the entire user's text to identify significant, meaningful phrases or sentences.
	\item Generate codes — labels that assign summative, salient, essence-capturing, and/or evocative attributes/meanings to text data. Coding helps organise data at a granular, specific level and reduces large amounts of data into small chunks of meaning. A code should be sufficiently well-defined and demarcated so that it does not overlap with other codes and should fit logically within a larger coding framework or template that guides the coding process by outlining and defining the codes to apply. Codes are semantic units corresponding to a specific section, part or even one paragraph of text. Be sure to identify and define each such code in the text. 
	\item Ensure each code is accompanied by a direct quotation from the user's text that exemplifies the code. 
\end{enumerate}
Output: Present a table with the following columns: \\
Column 1: Document name \\
Column 2: Code \\
Column 3: Quotation exemplifying the code from the user's text \\

\textbf{Step 2}: Code Clustering \\
Action Trigger: The user asks to proceed with Step 2. \\
Process: 
\begin{enumerate}
	\item Group up (via clustering) the generated codes into clusters based on similarity and thematic relatedness.
	\item Identify and articulate the abstract themes each cluster represents.
	\item For each cluster, maintain the specific codes and quotations from the user's text from Step 1.
	\item Analyse the frequency and co-occurrence of codes within each cluster to determine their relevance and prominence in the text.
\end{enumerate}
Output: Compile a table with the following columns: \\
Column 1: Document name \\
Column 2: Cluster \\
Column 3: Description of cluster meaning \\
Column 4: Code \\
Column 5: Quotation representing the code from the user's text \\

\textbf{Step 3}: Developing AI-Specific Codes \\
Action Trigger: The user asks to proceed with Step 3. \\
Process: 
\begin{enumerate}
	\item Scans the entire user's uploaded text to locate sections specifically discussing AI, its applications, ethical considerations, technological advancements, or any AI-related narratives.
	\item Within these identified sections, pinpoint key concepts, terms, arguments, and perspectives related to AI.
	\item Assign specific codes to these segments, where each code encapsulates a unique aspect of the AI narrative presented in the text. These codes should reflect the multifaceted nature of AI discussions, capturing technological, ethical, societal, and futuristic viewpoints.
	\item For each AI-specific code, provide a concise description that clarifies the aspect of AI the code represents, ensuring clarity and relevance to the AI discourse.
	\item Select precise quotations from the uploaded user's text that exemplify or best illustrate each AI-specific code.
\end{enumerate}
Output: Compile a table with the following columns: \\
Column 1: Document name \\
Column 2: AI-Specific Code \\
Column 3: Quotation representing the code from the user's text. 

This step hones in on the AI narrative within the user's text, ensuring that AI-related themes are thoroughly explored and accurately represented through specific coding. This focused approach allows for a deeper understanding and analysis of AI discussions, which can be pivotal in research within politics, international studies, and geopolitics, especially considering the growing influence of AI in these fields.

\textbf{Additional Instructions}:
\begin{enumerate}
	\item Accuracy and Relevance: Ensure that the analysis accurately represents the user's text's content and themes. Prioritise relevant information and themes pertinent to politics, international studies, and geopolitics.
	\item Depth of Analysis: Provide a thorough and nuanced analysis, offering deep insights into the user's text's themes and meanings.
	\item Scholarly Tone: Adopt a formal, scholarly tone when explaining concepts, methods, and findings.
	\item Avoid Personal Opinions: Maintain objectivity by avoiding personal opinions or interpretations not supported by the user's text.
	\item Clarification of Complex Concepts: Clearly explain any complex concepts or methodologies used in the analysis, providing examples where necessary.
\end{enumerate}
\textbf{User’s input:} 
\begin{itemize}
	\item Run Step 1 of the thematic analysis on this uploaded text. 
	\item Continuing to work with this text, run Step 2.
	\item Continuing to work with this text, run Step 3.
\end{itemize}

\printbibliography

\end{document}